\documentclass[twoside,a4paper,11pt]{sca}
\usepackage{graphicx}
\usepackage{hyperref}
\usepackage{natbib}  % Cross-reference package (Natural BiB)
\begin{document}
\pagenumbering{arabic}
\pagestyle{myheadings}
\thispagestyle{empty}
%%%%{\flushright\includegraphics[width=\textwidth,bb=58 650 590 680]{stamp.pdf}}
{\flushright\includegraphics[width=\textwidth,bb=90 650 520 700]{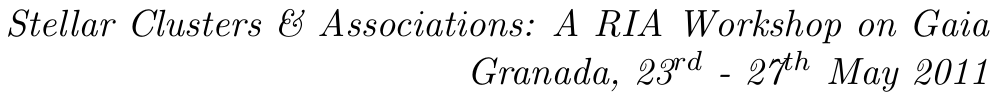}}
\vspace*{0.2cm}
\begin{flushleft}
{\bf {\LARGE
%
%%% TITLE of the paper. 
Large-scale young Gould Belt stars across Orion 
%
% Do not delete next few lines
}\\
\vspace*{1cm}
%
%%% Include here the LIST OF AUTHORS.
%%% Include here the LIST OF AUTHORS.
%%% Note that the last author has to be preceeded by an AND.
K. Biazzo$^{1}$,
J.M. Alcal\'a$^{1}$,
M.F. Sterzik$^{2}$,
E. Covino$^{1}$,
A. Frasca$^{3}$,
and 
P. Guillout$^{4}$
%
% Do not delete next few lines
}\\
\vspace*{0.5cm}
%
%%% AFFILIATIONS LIST.
%%% and the AFFILIATIONS LIST. Note that one affiliation per line.
%%% Add as many affiliations as necessary. 
$^{1}$
INAF - Capodimonte Astronomical Observatory\\
$^{2}$
European Southern Observatory (ESO) - Chile\\
$^{3}$
INAF - Catania Astrophysical Observatory\\
$^{4}$
Observatoire Astronomique de Strasbourg, CNRS, UMR 7550, France\\
%
% Do not delete next few lines
\end{flushleft}
%
% Headings
\markboth{
%%% Type the SHORT version of the paper title.
%%% Type the SHORT version of the paper title.
Large-scale young Gould Belt stars across Orion
}{ % Do not delete
%
%%%  First Author \& Second Author   OR   First-author et al. 
%%%  First Author \& Second Author   OR   First-author et al. if the author list 
%%% contains three or more authors.
K. Biazzo et al.
% 
% Do not delete next few lines
}
\thispagestyle{empty}
\vspace*{0.4cm}
\begin{minipage}[l]{0.09\textwidth}
\ 
\end{minipage}
\begin{minipage}[r]{0.9\textwidth}
\vspace{1cm}
\section*{Abstract}{\small
%
% ABSTRACT ABSTRACT ABSTRACT
% ABSTRACT ABSTRACT ABSTRACT
%%% Type the ABSTRACT of your paper
We report first results on the large-scale distribution of the ROSAT All-Sky Survey (RASS) 
X-ray sources in a 5000 deg$^2$ field centered on Orion. Our final aim is to study the properties of 
different widespread populations in the Orion Complex close to the Gould Belt (GB) in order to trace the star 
formation history in the solar neighbourhood.
%
% Do not delete next few lines
\normalsize}
\end{minipage}
%
%
%%% BODY of the paper
%%% BODY of the paper
%
\section{Sample definition and candidate selection}
We considered a $\sim$5000 deg$^2$ field centered on Orion and selected in this area $\sim$1500 young stellar object (YSO) candidates 
through X-ray criteria established by Sterzik et al. (1995). We then selected a $\sim$10$^{\rm o}$$\times$75$^{\rm o}$ strip (see Fig. 1) 
perpendicular to the GB and crossing the Orion star-forming region (SFR), as well as a $\sim$10$^{\rm o}$$\times$10$^{\rm o}$ region at 
$\alpha=5^{\rm h}34^{\rm m}$ and $\delta$=+22$^{\rm o}$01' with enhanced X-ray space density. Some $\sim$200 stars inside the strip turn 
out to be YSO candidates (Fig. 1), while three stellar groups seem to have high X-ray space density: two of them are inside the strip, 
while the third one is close to $\alpha=5^{\rm h}34^{\rm m}$ and $\delta$=+22$^{\rm o}$01'. 

\section{Observational data set and candidate characterization}
Low-resolution spectroscopy ($R\sim 1\,000$) was obtained with the Boller and Chivens Cassegrain spectrographs attached to the 1.5m 
telescope of the ESO (Chile) and to the 2.1m of the Observatorio Astron\'omico Nacional de San Pedro M\'artir 
(Mexico). High-resolution spectroscopy ($R\sim 30\,000-100\,000$) was performed using the FOCES spectrograph 
attached to the 2.2m telescope at the Calar Alto Observatory (Spain) and with the Coud\'e Ech\'elle Spectrometer 
fed by the 1.5m CAT telescope (Chile).

Using the low-resolution spectra, we determined spectral types (and effective temperatures) and detected the presence of H$\alpha$ and lithium 
line. This allowed us to select very young not-accreting stars (accretion is mainly associated with stars in the Orion SFR). We thus find 
that all stars located in the Orion region possess lithium stronger than the Pleiades stars of the same spectral type; many of the stars located 
in the region of the GB have also a strong Li content, but tend to be more similar to that of the Pleiades; and the majority of the field stars 
have lithium weaker than the Pleiades. From the high-resolution 
spectroscopy we determined rotational and radial velocity and measured lithium abundance. Preliminary results show that stars in the strip are 
segregated in three populations: clustered stars in Orion, with ages of 2-5 Myr; non-clustered stars in the GB with ages of 5-10 Myr; field 
stars with a wide age spread; while, stars in stellar groups show ages $<$20 Myr.

\begin{figure}
\center
\vspace{-1.cm}
\includegraphics[scale=0.45,angle=-90]{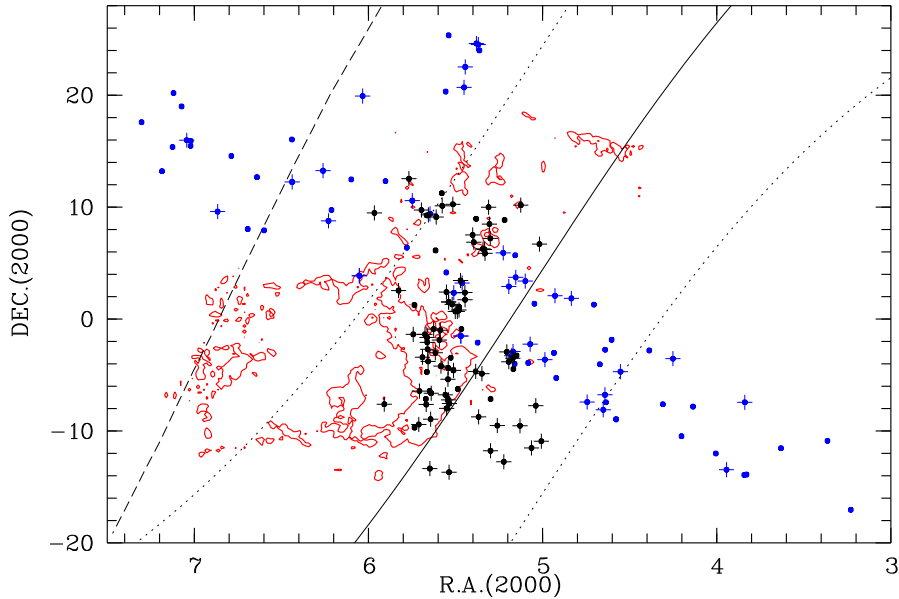} 
\vspace{-1cm}
\caption{\label{fig1} Large-scale spatial distribution of our targets in the strip and clumps observed at low and high resolution (blue symbols). 
The Alcal\'a et al. (2000) data are also shown with black symbols. Dots represent targets with low-Li content, while dots with crosses are 
stars with high-Li content. The positions of the three clumps are at ($\alpha=5^{\rm h}34^{\rm m}$, $\delta$=+22$^{\rm o}$01'), 
($\alpha=5^{\rm h}07^{\rm m}$, $\delta$=$-03^{\rm o}$20'), and ($\alpha=4^{\rm h}30^{\rm m}$, $\delta$=$-08^{\rm o}$). Solid and dotted lines 
represent the GB and its limits (Guillout et al. 1998), while the dashed line represents the Galactic Plane. The CO emission map by 
Maddalena et al. (1986) is also overlaid in red.
}
\end{figure}

\section{Work in progress and Future perspectives with GAIA}
Following the prescriptions given by Biazzo et al. (2011), we are measuring the metallicity of single stars with low $v\sin i$ and 
observed at high resolution. Two main preliminary results are the following: i) stars with high-Li content 
show a distribution in agreement with that of young nearby clusters, i.e. close to the solar one; ii) stars with no Li absorption show wide 
[Fe/H] values, which resemble the distribution of field stars in the solar neighborhood.

Thanks to the future GAIA mission we will obtain parallaxes for our targets with a precision of 0.014-0.070 mas, which will translate into 
1-7 pc at the Orion distance. This will allow us to derive the distance to the individual stars and to place them on the HR diagram 
with accuracy of 0.7-1.4\% in $\log(L/L\odot)$. Therefore, GAIA will allow us to definitively establish the nature of the widespread 
population of young stars on a Galactic scale.

%
%
% Do not delete the next line
\small  % Do not delete
%
%%% Comment the following line if you do not have acknowledgments.
%\section*{Acknowledgments}   % Do not delete if you declare acknowledgments
%
%%% ACKNOWLEDGMENTS
%%% ACKNOWLEDGMENTS
%If you do not have any acknowledgments, you may comment this Section.

%
% Do not delete the next few lines
%************************************************************************************%

%************************************************************************************%
%
% Do not delete the next few lines

\bibliographystyle{aa}
\bibliography{mnemonic,ref_user}

\end{document}